# A Sealed Gas Pixel Detector for X-ray Astronomy

R. Bellazzini[a*], G. Spandre[a], M. Minuti[a], L. Baldini[a], A. Brez[a], L. Latronico[a], N. Omodei[a], M. Razzano[a,b], M.M. Massai[a,b], M. Pinchera[a], M. Pesce-Rollins[a], C. Sgró[a], E. Costa[c], P. Soffitta[c], H. Sipila[d], E. Lempinen[d]

[a]*INFN sez.Pisa, Largo B. Pontecorvo, 3 I-56127 Pisa, Italy*

[b]*Dipartimento di Fisica ,Università di Pisa, Largo B. Pontecorvo, 3 I-56127 Pisa, Italy*

[c]*Istituto di Astrofsica Spaziale e Fisica Cosmica, Via del Fosso del Cavaliere 100, I-00133, Roma, Italy*

[d]*Oxford Instruments Analytical Oy, Nihtisillankuja 5 FI-02631 Espoo, Finland*

**Abstract**

We report on the results of a new, sealed, Gas Pixel Detector. The very compact design and the absence of the gas flow system, make this detector substantially ready for use as focal plane detector for future X-ray space telescopes. The instrument brings high sensitivity to X-ray polarimetry, which is the last unexplored field of X-ray astronomy. It derives the polarization information from the track of the photoelectrons that are imaged by a high gain (>1000), fine pitch GEM that matches the pitch of a pixel ASIC which is the collecting anode of the GPD (105k, 50μm wide, hexagonal cells). The device is able to simultaneously perform good imaging (50÷60μm), moderate spectroscopy (~15% at 6 keV) as well as fast, high rate timing in the 1-10keV range. Moreover, being truly 2D, it is non dispersive and does not require any rotation. The great improvement of sensitivity, at least two orders of magnitude with respect to traditional polarimeters (based on Bragg crystals or Thomson scattering), will allow the direct exploration of the most dramatic objects of the X-ray sky. At the focus of the large mirror area of the XEUS telescope it will be decisive in reaching many of the scientific goals of the mission. With integration times of the order of one day, polarimetry of Active Galactic Nuclei at the per cent level will be possible, making for a real breakthrough in high energy astrophysics.

## 1. Introduction

We have built and tested a sealed Gas Pixel Detector (GPD). The detector was filled with the standard gas mixture optimized for X-ray polarimetry in the 1-10 keV range (Ne-DME 50%-50%) and then sealed and operated for more than one month without any gas refilling. Low outgassing materials and adhesives were used and special mounting procedures were taken to allow a long term stable behavior of the detector. The sealed GPD is a detector with a level of integration, compactness and operational simplicity comparable to solid state detectors. To reach this goal, since the year 2001 we have been actively working within the INFN funded PIXI project on the GPD concept in which a custom CMOS analog chip is simultaneously the pixelized charge collecting electrode and the front-end electronics of a suitable charge multiplier, a single GEM foil in our case. GPDs equipped with PIXI ASICs of three different generations (see fig.1) have been designed and fabricated [1,2,3]. The three generations are characterized by increased size, reduced pitch and improved functionality. The sealed GPD is based on the third generation ASIC (0.18 μm CMOS technology, 105600 pixels at 50 μm pitch organized in a honeycomb array).

* Corresponding author. Tel.: +30-050-2214-367; fax: +39-050-2214-317; e-mail: ronaldo.bellazzini@pi.infn.it.

The detector provides unique polarimetric performance but also remarkable imaging, spectral and timing capabilities and for these reasons it has been proposed as photoelectric polarimeter at the focus of a JET-X like optics in a Small Satellite mission (POLARIX, [4]) and at the focus of the large area mirror of the XEUS telescope [5], the ESA permanent space borne X-ray Observatory planned to be launched in 2015. The sensitivity and efficiency of the instrument onboard of the XEUS observatory will allow to perform energy and angular resolved polarimetry at the level of few % on many galactic and extragalactic sources with photon fluxes down to milliCrab or fractions.

## 2. The Gas Pixel Detector as X-Ray Polarimeter

Schematically a Gas Pixel Detector (GPD) for X-ray polarimetry is made by a gas cell with a Berillium window, a Gas Electron Multiplier (GEM) and a pixellated charge collection plane, which is directly connected to the analog readout electronics. The GEM amplifies the charge of the electron tracks generated in the drift gap and therefore providing the energy and time information.

In the photoelectric absorption of polarized X-rays, a “$\cos^2$” modulation is observed with respect to the azimuthal angle φ between the direction of emission of the photoelectron and the polarization vector. With modern gas pixel detectors it is possible to image, with high accuracy, the single photoelectron track and to derive the conversion point, which differs from the barycenter. Consequently, it is possible to determine the initial direction of the photoelectron path before it is randomized by the collisions with the gas molecules. The angle and degree of polarization of the X-ray photon are then obtained from the distribution of the reconstructed emission angles. Although, it is worth noticing that the isotropically emitted Auger electron is not modulated by the X-ray polarization and therefore represents a disturbance, especially at low energy.

The top metal layer of the readout ASIC, which is patterned in a honeycomb hexagonal pattern, is the readout plane of the charge produced by the photoelectrons. Each pixel is connected to a full electronics chain (pre-amplifier, shaping amplifier, sample and hold, multiplexer) built immediately below it, realized exploiting the remaining metal, polysilicon and oxide layers of the deep sub-micron VLSI CMOS technology.

The third generation PIXI chip has 105600 hexagonal pixels arranged at 50μm pitch in a 300×352 honeycomb matrix, corresponding to an active area of $15\times15\text{mm}^2$ with a pixel density of $470/\text{mm}^2$. The chip integrates more than 16.5 million transistors and it is subdivided in 16 identical clusters of 6600 pixels (22 rows of 300 pixels) or alternatively in 8 clusters of 13200 pixels (44 rows of 300 pixels) each one with an independent differential analog output buffer.

Each cluster has a customizable internal self-triggering capability with independently adjustable thresholds. Every 4 pixels (mini-cluster, fig.2) contribute to a local trigger with a dedicated amplifier whose shaping time (Tshaping~1.5 μs) is roughly a factor of two faster than the shaping time of the analog charge signal. The contribution of any pixel to the trigger can be disabled by direct addressing of the pixel. An internal wired-OR combination of each mini-cluster self-triggering circuit holds the maximum of the shaped signal on each pixel. The event is localized in a rectangular area containing all triggered miniclusters plus a user selectable margin of 10 or 20 pixels. The Xmin, Xmax and Ymin, Ymax rectangle coordinates are available as four 9-bit data outputs as soon as the data acquisition process following an internally triggered event has terminated, flagged by the DataReady output. The event window coordinates can be copied into a Serial-Parallel IO interface register (a 36-stage FIFO) by applying an external command signal (ReadMinMax). Subsequently, clock pulses push out the analog data to a serial balanced output buffer compatible with the input stage of the Texas Instruments 12 bit flash ADC ADS572x.

In self-trigger operation the read-out time and the amount of data to be transferred result vastly reduced (at least by a factor 100) with respect to the standard sequential read-out mode of the full matrix (which is still available, anyway). This is due to the relatively small number of pixels (600-700) within the region of interest. The self-trigger output signal from the chip can also be used to derive timing information. Given the relatively slow shaping time of the trigger amplifier, a resolution of the order of 1 μs can be expected. Better results (of the order of 100ns) could be obtained, if needed, by using the much faster GEM signal.

The main characteristics of the chip in comparison with the well known Medipix2 chip [6] are summarized in Table 1.
A custom and very compact DAQ system to generate and handle command signals to/from the chip (implemented on Altera FPGA Cyclone EP1C240), to read and digitally convert the analog data (ADS5270TI Flash ADC) and to store them, temporarily, on a static RAM, has been developed. By using the RISC processor NIOS II, embedded on Altera FPGA, and the self-triggering functionality of the chip, it is possible to acquire the pedestals of the pixels in the same chip-defined event window (region of interest) immediately after the event is read-out. The readout of the pedestals is user-defined and can be performed once as well as several times.
The average of the pedestal readings is used to transfer pedestal subtracted data to the off-line analysis system. This mode of operation has the great advantage of allowing the real time control of the data quality and to cancel any effect of time drift of the pedestal values or other temperature or environmental effects.
The slight disadvantages are a small increase of the channel noise (at maximum a factor √2, for the case of 1 pedestal reading only) and an increase of the event read-out time. For most of the applications we envisage, this is not a real problem given the very large signal to noise ratio (well above 100) and the very fast window mode operation. Nevertheless, the standard mode of operation with the acquisition of a set of pedestal values for all the 105k channels at the beginning or at the end of a data taking run is still possible.
The instrument control and data acquisition is done through a VI graphic interface developed in LAbVIEW. The VI performs a bidirectional communication through a 100Mbps TCP connection between the DAQ and a portable PC running Windows XP.

## 3. The construction of the sealed detector

The assembly of the sealed detector was done in Finland in collaboration with Oxford Instruments Analytical Oy, a small company with long lasting experience in sealed and space qualified instruments.

The detector was built up directly over the chip case (fig.3). As a first step, a 50μm thick Gas Electron Multiplier with 50μm pitch holes on a triangular pattern is mounted on top of the chip (fig.4). The collection gap between the bottom GEM plane and the pixel matrix is only 1 mm thick. The photon absorption region between the top GEM plane and the enclosure drift electrode (50μm Be window), has been realized with a 10 mm thick ceramic spacer. At the end of the assembly process the detector was vacuum pumped and heated for several days and then filled with our standard gas mixtures: 50% Neon – 50% DME. The device was sealed by crimping the small copper tube inlet and it was then sent to INFN-Pisa (Italy) for testing. A photo of this light (80gr, including the small motherboard), compact, 'vacuum valve' type assembly is shown in fig5 (HV protection box removed).

## 4. Performance of the sealed GPD

At present we have accumulated 38 days of gas gain stability monitoring.  Even though our original target of 1 month survival has been largely achieved, the measurement is still on-going without any sign of degradation. The high voltage settings used in the test are: VDRIFT = -2000V, VGEM(top) = -750V, VGEM(bottom) = -300V, the read-out electrode being at ~0V (the charge preamp input voltage). We used a $^{55}$Fe source with a 450Hz continuous detection rate. This rate corresponds to the observation of the Crab with a 800 cm2 effective area mirror. The average signal amplitude from the upper electrode of the GEM was registered periodically (2-3 times/day) using the on-line measurement capability of a digital oscilloscope. The temperature and the pressure of the laboratory were also recorded. Overall, the behavior of the detector was very stable(see fig.6). A day-night modulation anti-correlated with temperature fluctuations is learly detectable. No correlation was found with the atmospheric pressure. The origin of the small (13%) steady gain increase in the first three weeks of operation is under study. Fig.7 shows the GEM signals and their pulse height distribution at the end of the 5 weeks observation period as recorded directly on the digital oscilloscope..
The 50μm pitch GEM has shown to work magnificently. It has a large effective gain (well above 1000) at a much reduced voltage (at least 70 Volt less) compared with our previous 90μm pitch GEM. This is likely due to the higher field lines density inside the very narrow amplification holes.

The technological challenge in the fabrication of this type of GEM was the precise and uniform etching of these narrow charge multiplication holes (33 and 15μm diameter at the top and in the middle of the kapton layer) on a GEM foil of standard 50μm thickness. The use of a GEM with a much finer pitch than usual (pitch and thickness have now equal size) and well matching the 50μm read-out pitch has pushed forward the 2D imaging capability of the device, and therefore allowing to reach a very high degree of detail in the photo-electron track reconstruction (see fig.8). This is of great importance for our X-Ray Polarimetry application, especially when working at low photon energy (2-3 keV) where the tracks are very short. The Monte Carlo prediction of the position resolution of the photon conversion point reconstruction agrees with our experimental tests with phantoms and is in the 50-60μm range [3]. Obviously, the fine grain reconstruction of the initial part of the photo-electron track allows a better estimation of the emission direction and from the obtained angular distribution a better evaluation of the degree and angle of polarization of the detected radiation.

## 5. Polarimetric sensitivity

The measurement of the modulation factor for polarized photons has been carried out by using radiation from a Cr X-ray tube (5.41 keV line). The X-ray beam is Thomson scattered through a Li target (6mm in diameter, 70mm long), canned in a beryllium case (500 μm thick) in order to prevent oxidation and nitridation from air[ 7]. The geometry of the output window of the scatterer and the distance with respect to the detector, limits the scattering angles to ~90º so that the radiation impinging the detector is highly linearly polarized (better than 98%). A modulation factor of 51.11%±0.89 has been measured (see fig.9). The GPD described in this paper is not only an excellent imager but also a good proportional counter with energy resolution of about 15% at 5.9 keV [8]. This feature is suitable for crucial energy resolved measurements, p.e. in polarimetry to separate the unpolarized fluorescence lines from the partially polarized continuum in reflection spectra. This property will allow for example to test General Relativity effects in the matter around a Black Hole through the measurement of the polarization angle as a function of energy [10,11].
All the physics processes ruling the operation of this detector as an X-ray polarimeter have been completely Monte Carlo simulated. These processes include the photoelectric interaction, the scattering and slowing of the primary electrons in the gas, drift and diffusion, gas multiplication and the final charge collection on the readout plane. All of which are functions of the photon energy and of the gas parameters. Description of the model can be found elsewhere [9]. Based on this MC, an extensive study of the fundamental parameters has been carried out. These parameters include the modulation factor and the detection efficiency, that together with the mirror effective area, the observation time and the celestial source flux, determine the sensitivity of the polarimeter. The resulting Minimum Detectable Polarization (MDP) is plotted in fig.10, assuming the actual polarimeter at the focus of the optics of XEUS, for different exposure times and for a few representative sources. With observations of one day we can measure the polarization of several AGNs down to few % level. Because of this high sensitivity the detector has been proposed as a focal plane instrument of a large area telescope such as those foreseen for the New Generation X-ray Telescope in the frame of the ESA Cosmic Vision 2015-2025.

## 6. Conclusions

We have demonstrated the stable operation of a very compact, sealed, GPD. The detector is now operating on a compact readout board communicating via Ethernet or a standard USB port with a normal PC without the need of a gas system. With devices like the one described in this paper the class of Gas Pixel Detectors has reached a level of operational reliability, integration, compactness and resolving power so far considered only to be in the reach of solid state detectors. As for the X-Ray Polarimetry application, the very low residual modulation and a modulation factor well above 50% will allow polarimetric measurements at the level of ~1% for hundreds of galactic and extragalactic sources: a real breakthrough in X-ray astronomy.

## Figures and tables

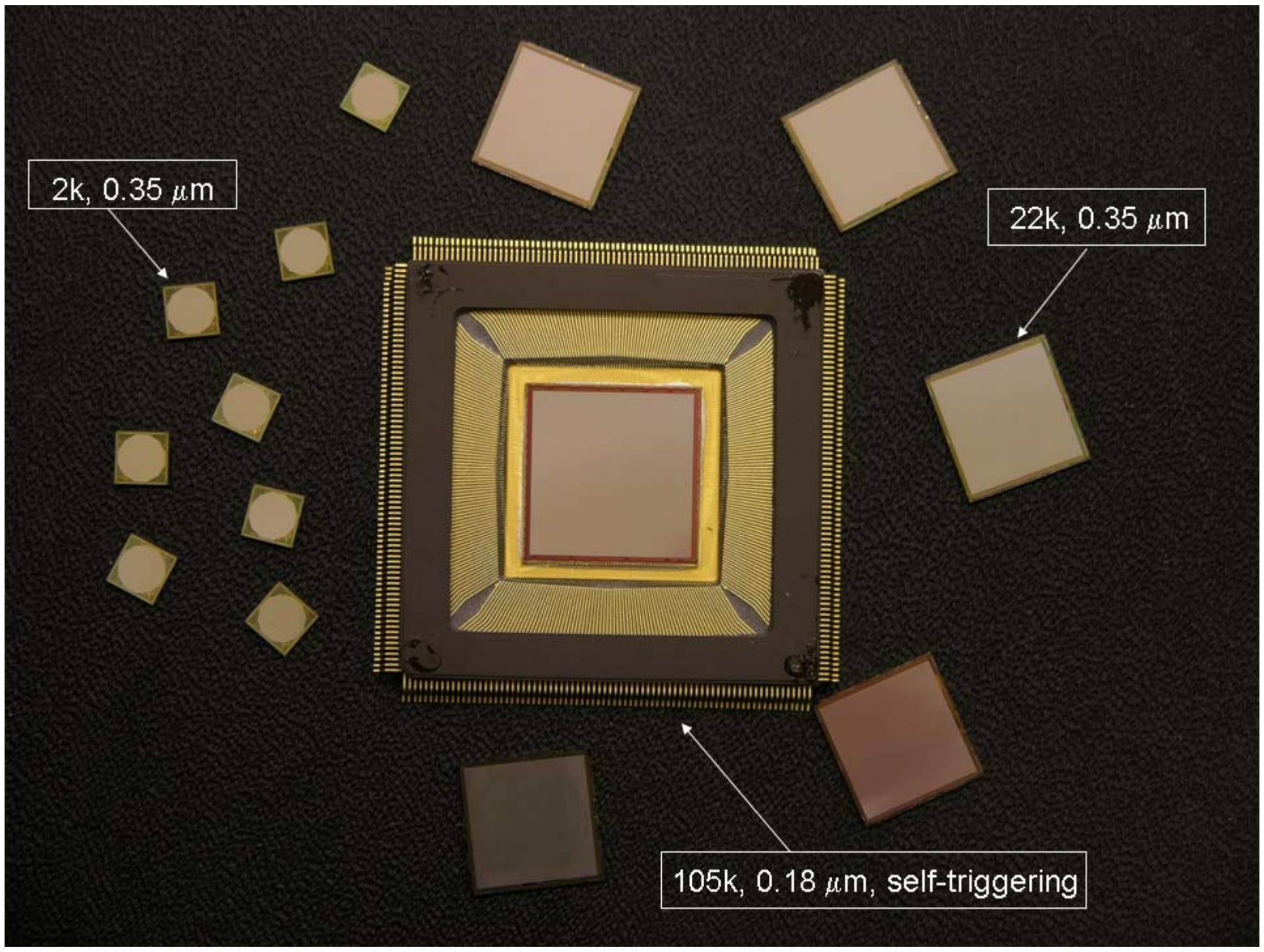

Fig. 1- The three PIXI chip generations in comparison.
The last version with 105.600 pixels is shown bonded to its ceramic package (304 pins).

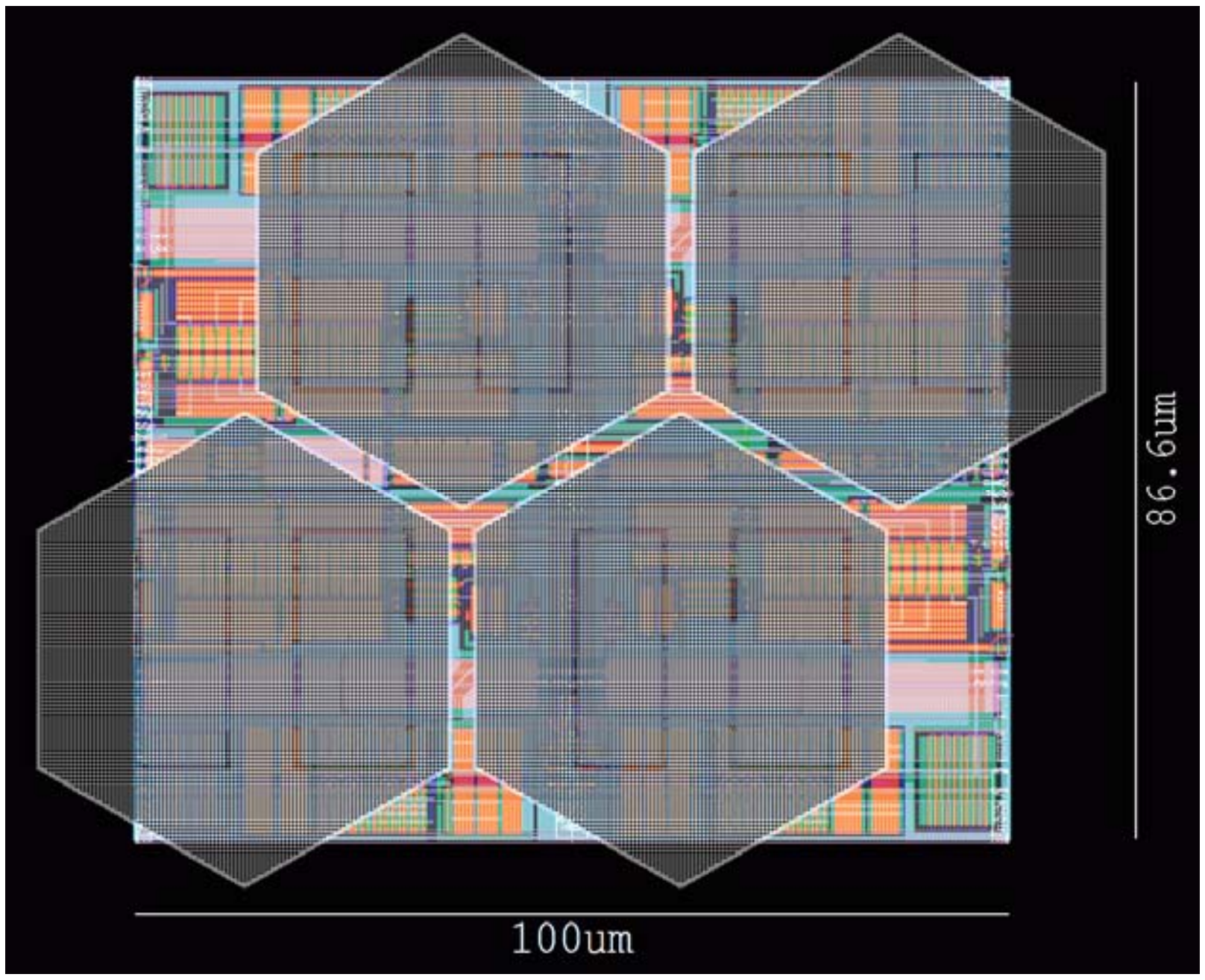

Fig. 2 - The 4 pixel self-trigger mini-cluster definition.

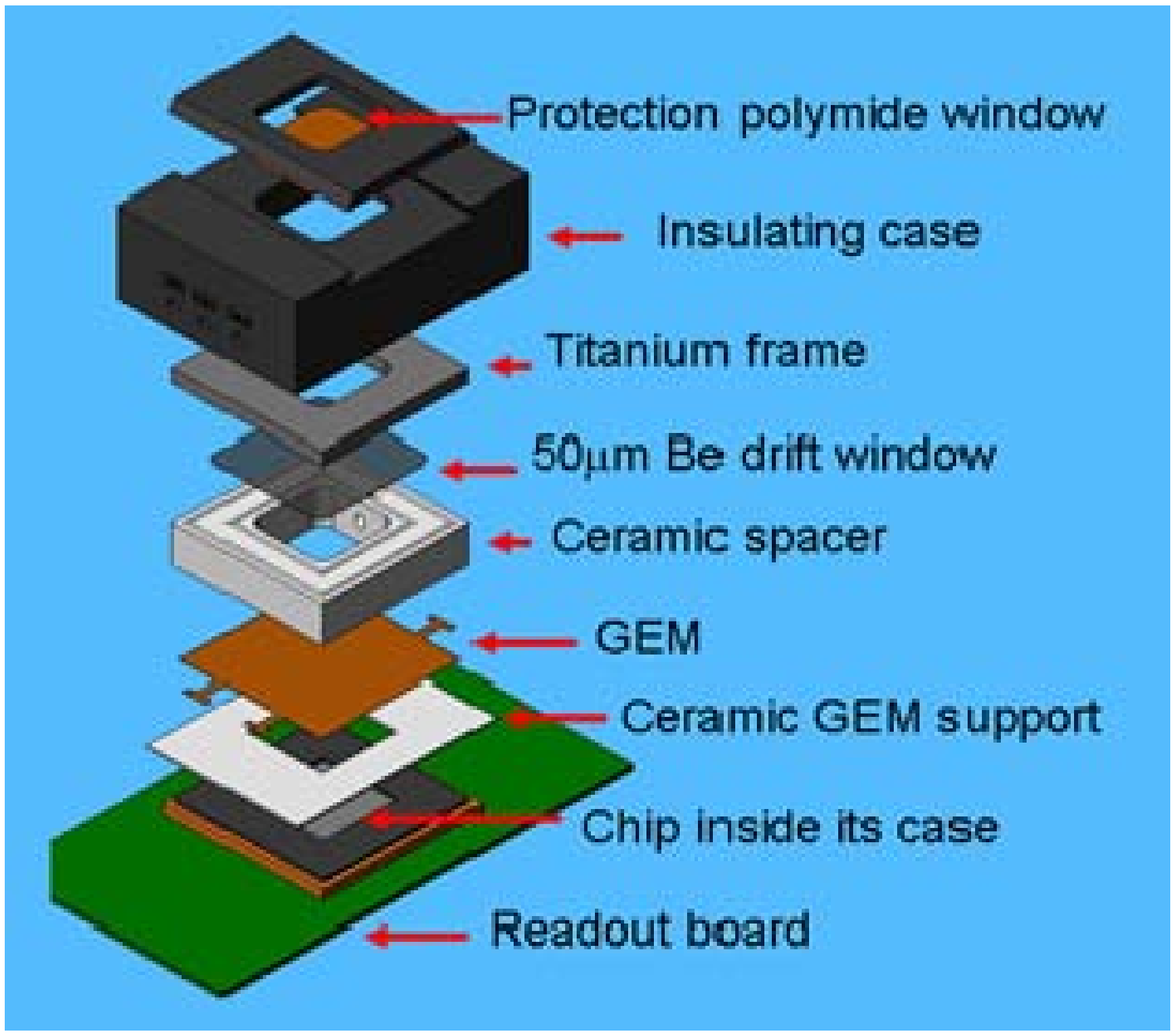

Fig. 3 - Exploded view of the GPD.

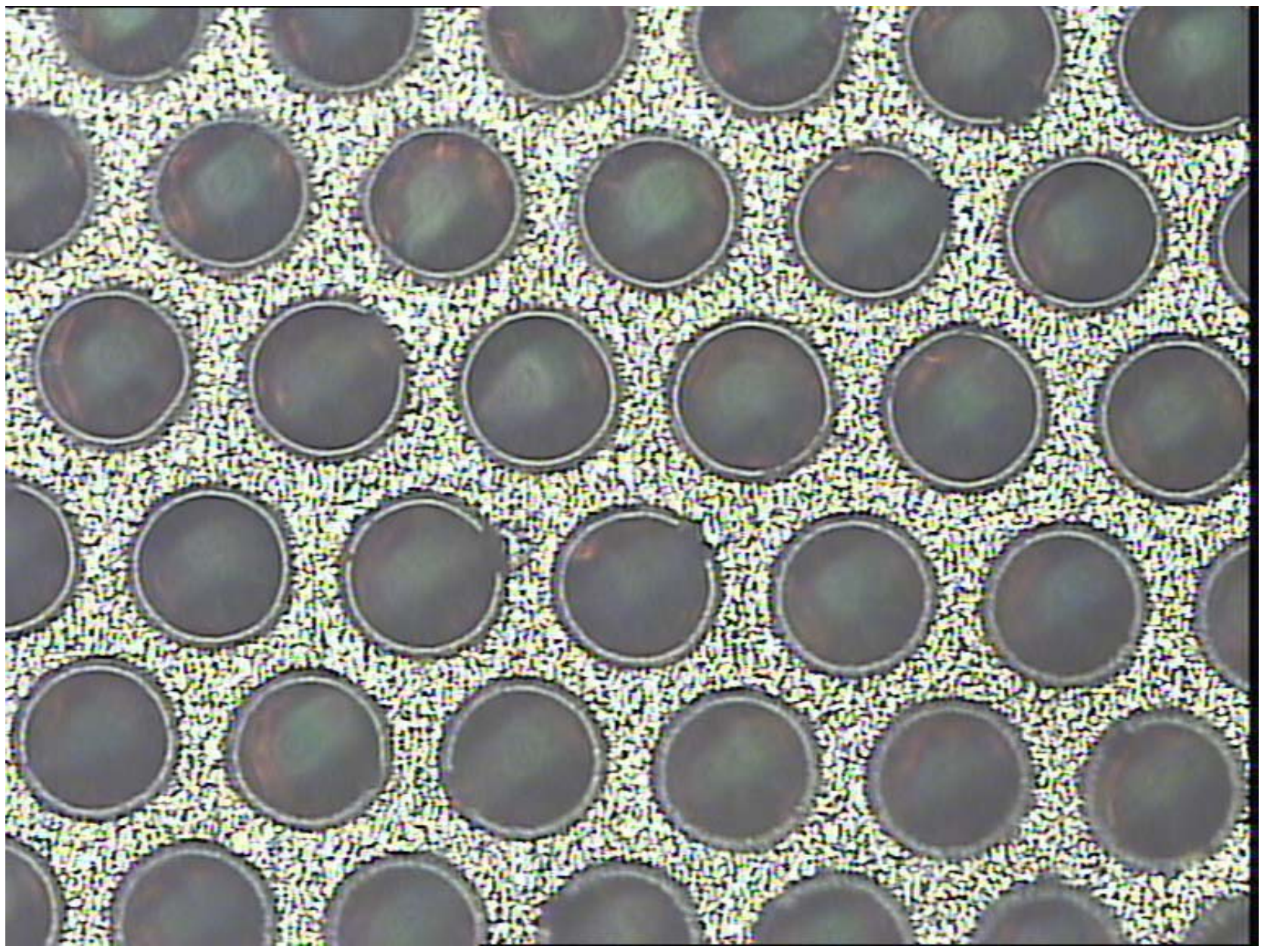

Fig. 4 – Microscope picture of the top GEM layer with 50μm pitch holes.

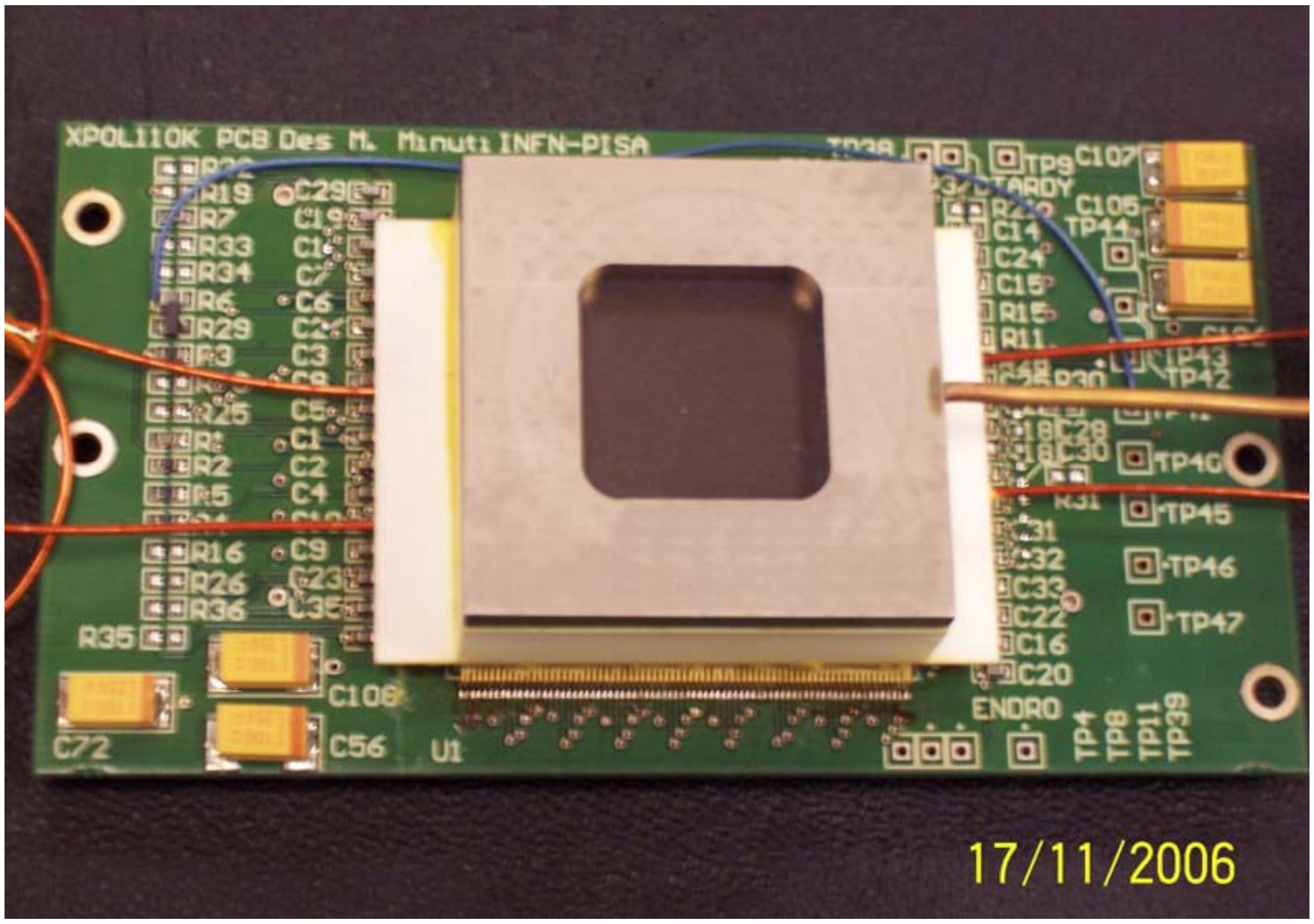

Fig. 5 - The sealed GPD.The 50 micron thick Be window brazed to the titanium frame is visible

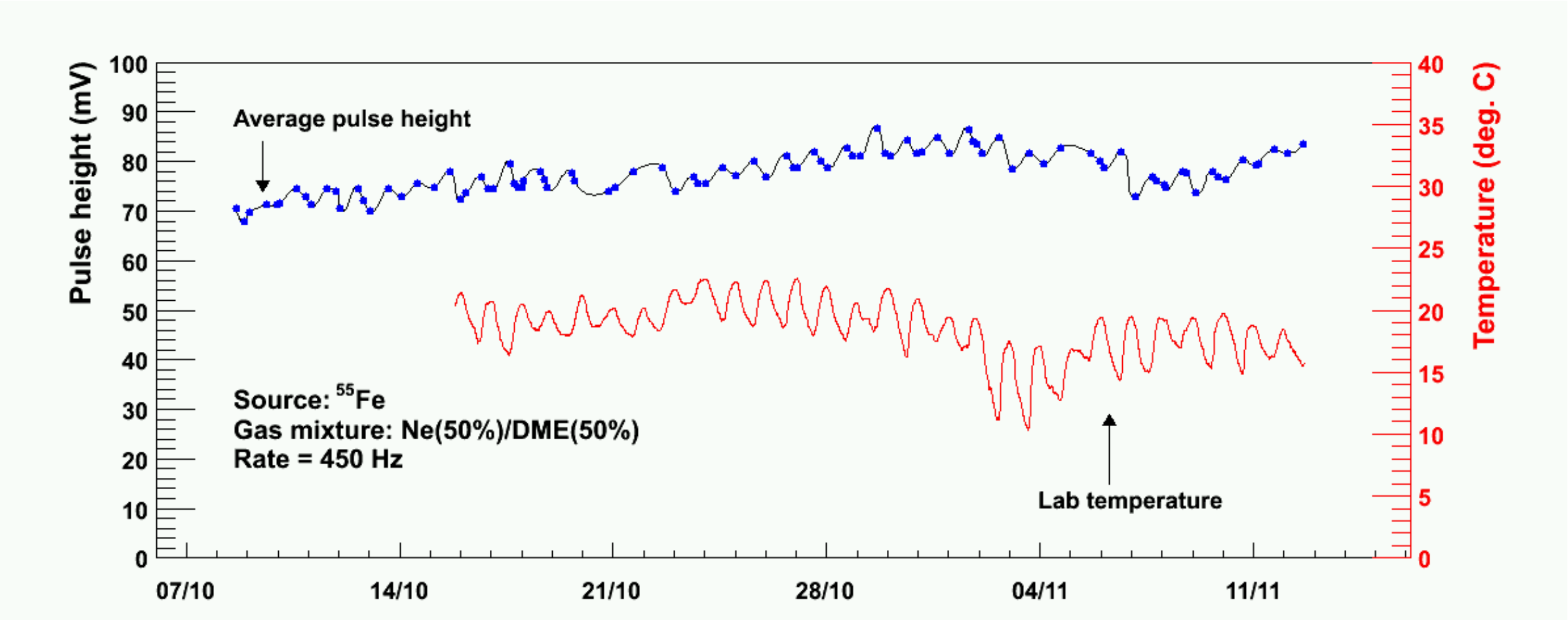

Fig. 6 - Monitoring the pulse height stability of the signal taken from the upper face of the GEM foil . Source:$^{55}$Fe, rate: 450Hz. In red the laboratory temperature.

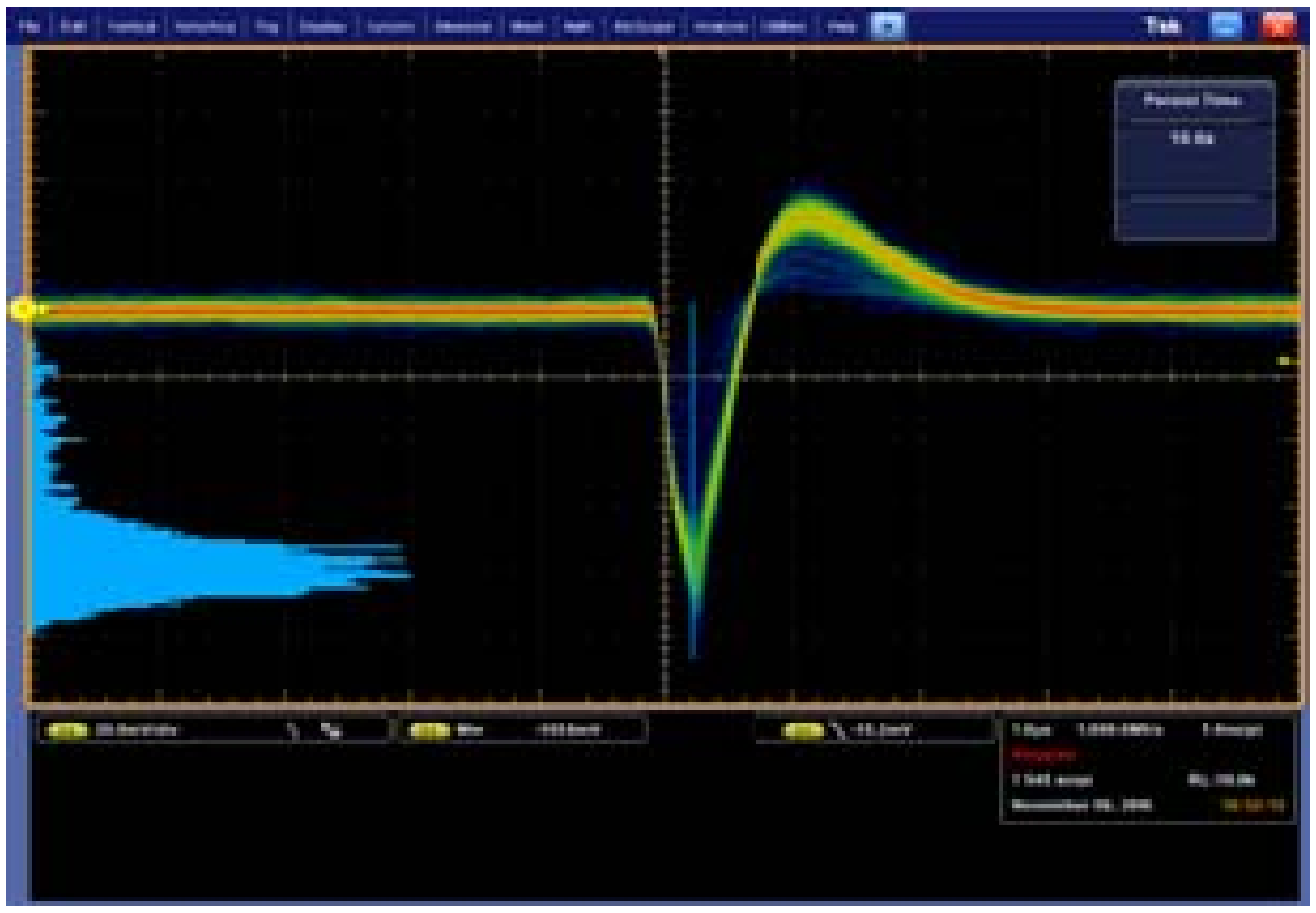

Fig. 7 - The pulses from the upper face of the GEM foil after more than one month of continuous operation..

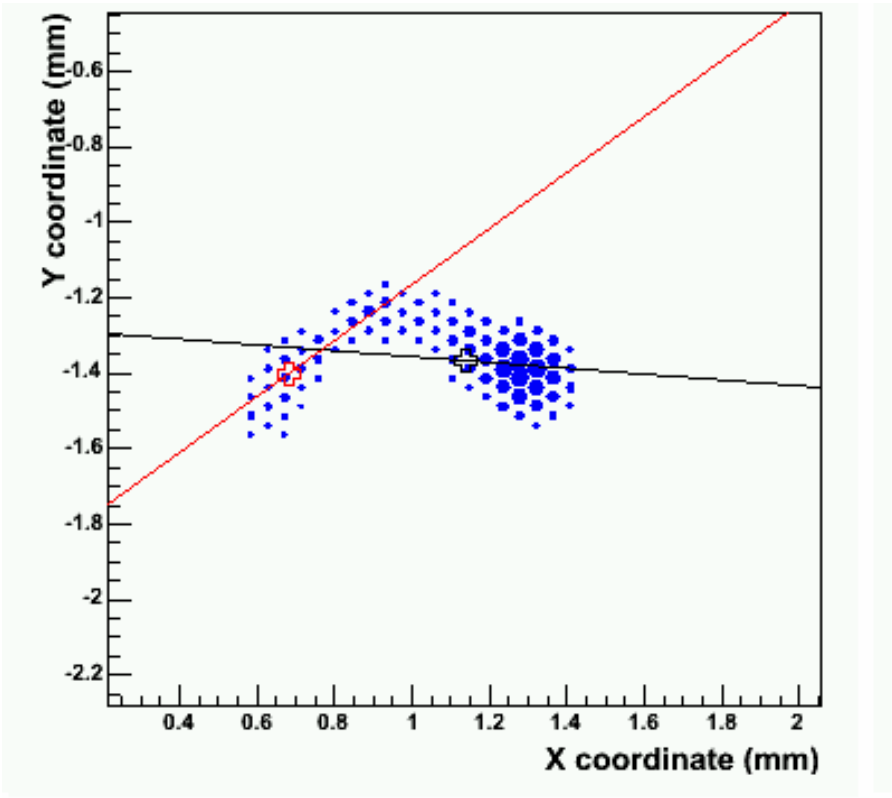

Fig.8 – 2D reconstruction of a *real* 5.9 keV photo-electron track.

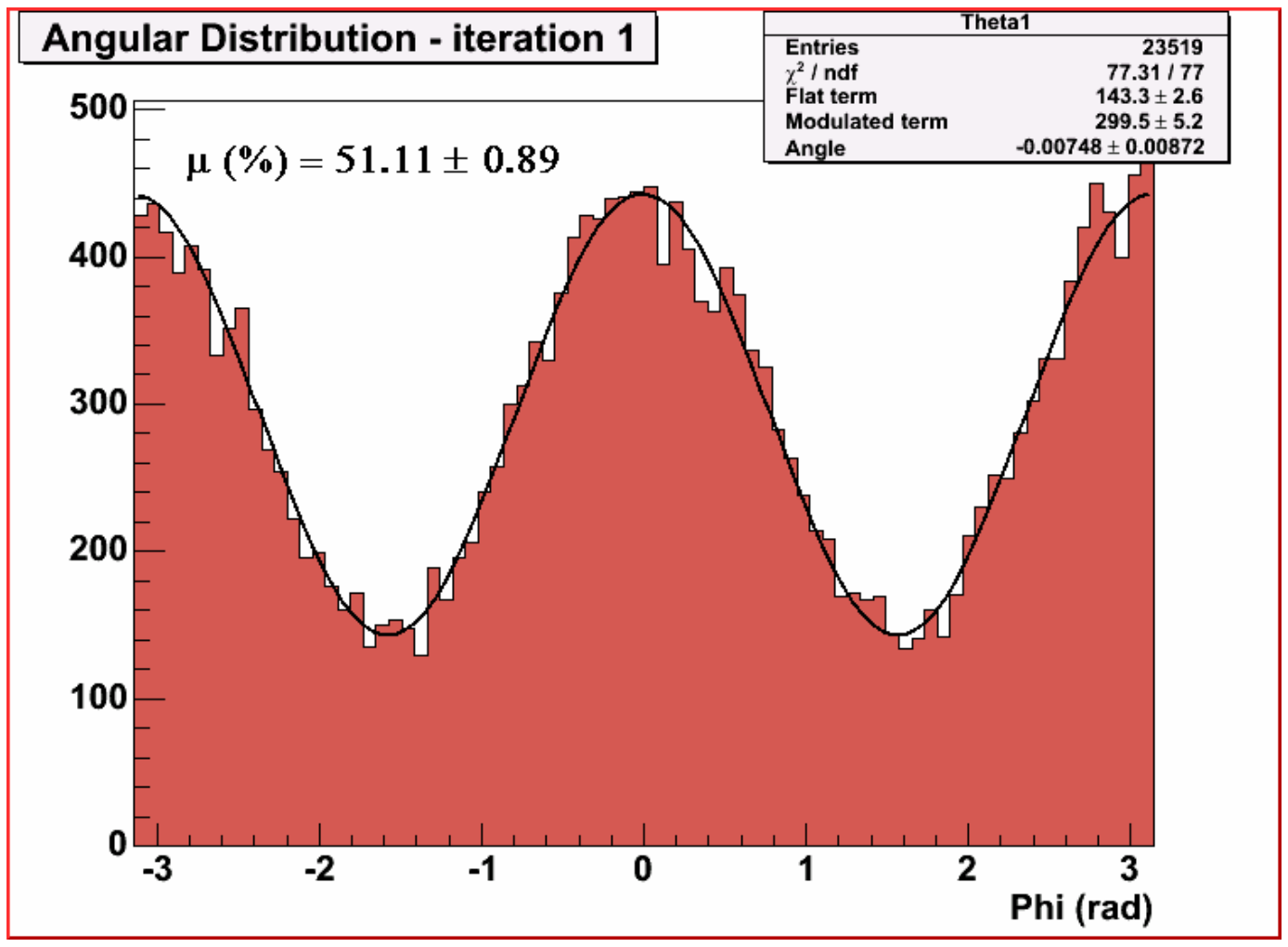

Fig. 9 - The angular distribution of 5.41 keV photoelectrons from a fully polarized beam

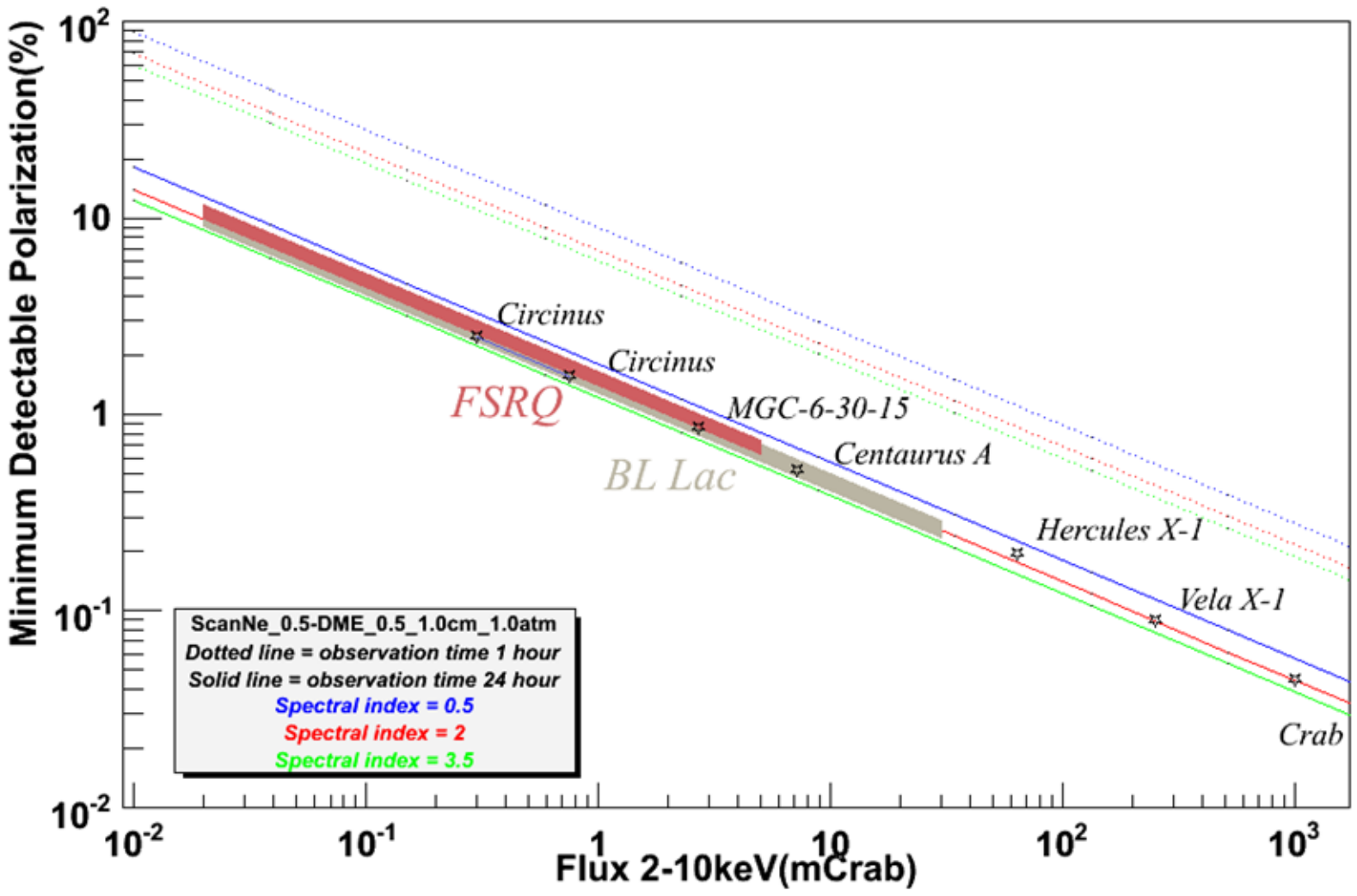

Fig. 10 - Minimum Detectable Polarization (MPD) for few representative sources using the GPD in the focal plane of XEUS

Table 1

Comparison of the main performance of the PIXI chip with the Medipix2 chip [6]

| | PIXI | Medipix2 |
|---|---|---|
| **Technology** | CMOS 0.18μm | CMOS 0.25μm |
| **Type** | analog | Digital (counting) |
| **Area** | 15×15 mm$^2$ | 14×14 mm$^2$ |
| **Pixel no.** | 105,600 | 65,536 |
| **Pixel density** | 470/mm$^2$ | 330/mm$^2$ |
| **Pixel noise** | 50 el. ENC | 110 el. ENC |
| **Readout scheme** | Asynchronous, synchronous | synchronous |
| **Readout trigger** | Self-trigger, internal, external | Internal, external |
| **Readout mode** | Single pixel, window,<br>full frame (8-16 nodes) | Full frame (1 node) |
| **Global threshold** | 2000 el. (adjusted) | 1000 el. (adjusted) |
| **Frame rate** | 10 KHz | 1 KHz |
| **Event rate** | $\sim 10^5$/s($10^2$ev./frame) | $\sim 10^9$/s |
| **Resolution** | ~4μm (analog interpolation) | ~15μm (55/√12) |